\documentclass[prl,twocolumn,showpacs,amsmath,amssymb]{revtex4}

\usepackage{graphicx}
\usepackage{dcolumn}
\usepackage{bm}

\def\tmo{TbMnO$_3$}

\def\be{\begin{equation}}
\def\ee{\end{equation}}

\begin{document}

\title{First-principles study of improper ferroelectricity in \tmo}

\author{Andrei Malashevich}
 \email{andreim@physics.rutgers.edu}
\author{David Vanderbilt}
\affiliation{
Department of Physics \& Astronomy, Rutgers University,
Piscataway, NJ 08854-8019, USA
}

\date{\today}

\begin{abstract}
We have carried out a first-principles theoretical study of the
magnetically induced polarization in orthorhombic \tmo, a prototypical
material in which a cycloidal spin structure generates an
electric polarization via the spin-orbit interaction.  We compute
both the electronic and the lattice-mediated contributions
to the polarization and find that the latter is strongly dominant.
We analyze the spin-orbit induced forces and lattice displacements
from both atomic and mode-decomposition viewpoints,
and show that a simple model based on nearest Mn--Mn neighbor
Dzyaloshinskii-Moriya interactions is not able to account fully
for the results.  The direction and magnitude of our computed polarization
are in good agreement with experiment.
\end{abstract}


\maketitle

Multiferroic and magnetoelectric materials are currently under
intensive study, both from the point of view of fundamental
materials physics and because of their potential application in
novel technological devices such as memories, sensors, and
transducers \cite{fiebig}.  An interesting class of systems are
``improper ferroelectrics''
in which the magnetic order induces an electric polarization
in a material that is otherwise structurally centrosymmetric.  This
can happen in collinear spin systems if the spin pattern
breaks inversion symmetry; examples include
$R$Mn$_2$O$_5$ ($R$=Tb, Y, etc.) \cite{chapon:2004,chapon:2006},
the E phase of orthorhombic $R$MnO$_3$ ($R$=Ho, Tb)
\cite{munoz:2001,Sergienko:2006a}, \tmo\ in magnetic field
\cite{aliouane:2006}, and Ca$_3$Co$_{2-x}$Mn$_x$O$_6$
\cite{choi:2008}.
In such cases, symmetric exchange interactions can
explain the development of polar order, i.e., via exchange
striction effects \cite{Sergienko:2006a,cheong-mostovoy}.
Here we shall be concerned with another class of systems in which
a non-collinear cycloidal spin structure induces an electric
polarization via the spin-orbit (SO) interaction
\cite{Kimura:2007}; examples include Ni$_3$V$_2$O$_8$ \cite{lawes:2005},
CuFeO$_2$ \cite{Kimura:2006}, and orthorhombic $R$MnO$_3$
($R$=Tb, Dy, Ho, etc.) \cite{kimura,Kenzelmann:2005,Yamasaki-both}.
While such spiral spin structures and the resulting polarizations
usually appear only at low temperature, these systems are of special
interest because the coupling between magnetic and electric degrees
of freedom is so profound.

Theoretically, symmetry analysis and phenomenological models
have clarified the circumstances under which a spiral spin structure
can give rise to an electric polarization
\cite{Mostovoy:2006,Radaelli:2007,harris},
but are not well suited to identifying the dominant microscopic
mechanism responsible for the polar order.  Microscopic models
have been introduced and used to discuss possible mechanisms, which
can be divided broadly into two categories.  The first are purely
electronic mechanisms, in which the SO interaction (SOI) modifies the
hybridization of electronic orbitals in such a way as to shift the
center of charge
\cite{Katsura:2005,Bruno:2005,jia:2006,jia:2007}.
The second are ``lattice mechanisms'' involving magnetically-induced
ionic displacements, usually discussed in terms of Dzyaloshinskii-Moriya
(DM) interactions \cite{Sergienko:2006,harris,Hu:2007}.
However, it is generally difficult to estimate the magnitudes (and
even signs) of these two kinds of contributions from the models.
Because experiments have not been sensitive enough to resolve the
tiny SO-induced atomic displacements, a central unanswered question is
whether electronic or lattice mechanisms are dominant.
It is also unclear whether the
pattern of displacements should follow the predictions of
a simple DM model, or whether more complicated interactions enter
the picture.

First-principles calculations can play an invaluable role in resolving
such questions.
Density-functional theory (DFT) has proven to be an extremely useful
tool in understanding the physics of ordinary ferroelectric materials,
and has already been used to make important contributions to the
understanding of improper magnetic ferroelectrics of the collinear-spin
type \cite{hill_m,wang,picozzi:2007} and in the spiral magnetic materials LiCu$_2$O$_2$
and LiCuVO$_4$ \cite{xiang_whangbo}.  However, to our knowledge such methods
have not previously been applied to study the cycloidal magnetic states
in the orthorhombic perovskite system.

In this Letter, we choose orthorhombic \tmo\ as a paradigmatic
system of the cycloidal type, and carry out a detailed
first-principles study of the mechanisms by which the cycloidal
magnetic order gives rise to the polarization.  We demonstrate
that DFT calculations predict the appearance of electric polarization
in the cycloidal state via two mechanisms, a purely electronic one
and a lattice-mediated one, and that the latter is strongly dominant.
Nevertheless, we find that the DM theory alone cannot explain
the computed pattern of atomic displacements, suggesting that
other interactions not included in simple DM models play an
important role.

Ferroelectricity coupled to spiral magnetic order was discovered in
the orthorhombic perovskite \tmo\ (space group {\it Pbnm})
\cite{kimura} with a 20-atom cell containing four formula units.
The Mn$^{3+}$ ions form a collinear sinusoidal spin wave at
temperatures between $\sim$27\,K and $\sim$41\,K, but below
$\sim$27\,K a cycloidal spin wave forms with incommensurate
wavevector $k_{\rm s}\sim0.28$ along $b$, and a polarization
simultaneously appears along $c$ \cite{kimura,Kimura:2007}.

\begin{figure}
\includegraphics[height=1.7in]{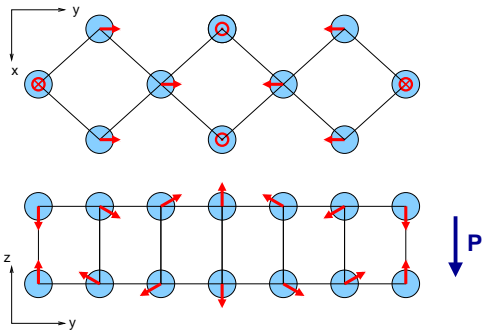}
\caption{\label{fig:rums}(Color online)
Sketch of $a\times3b\times c/2$ orthorhombic cell of \tmo\ ({\it Pbnm}
space group) showing Mn$^{3+}$ sites with a cycloidal magnetic structure.}
\end{figure}

In this work, electronic-structure calculations are
carried out using a plane-wave pseudopotential approach to
DFT as implemented within the VASP code \cite{kresse-vasp}
using PAW potentials\cite{Blochl,Kresse:1999}.
(The Tb potential does not have $f$ electrons in the valence.)
We use the local-density approximation (Ceperley-Alder\cite{Ceperley}
with Vosko-Wilk-Nusair correlation\cite{voskown}) 
with on-site Coulomb interactions
(LDA+U). We let $U$=1\,eV to make the band gap close to the experimental
value of $\sim0.5$\,eV \cite{Cui:2005}.
Since the first-principles calculations require periodic boundary
conditions, we study a system in which the cycloidal
spin structure is commensurate with $k_{\rm s}=1/3$, making a supercell
that is tripled in $b$-direction (60 atoms per cell) as shown in
Fig.\,\ref{fig:rums}.  This is close enough to the
experimental $k_{\rm s}$ of 0.28 that our results can be expected
to be qualitatively and semiquantitatively meaningful.  A
3$\times$1$\times$2 $k$-point sampling is used.
The plane-wave energy cutoff was 500 eV, and
the electric polarization was calculated using the Berry-phase
approach \cite{King-Smith}. 

A structural relaxation was performed for the 60-atom supercell
without SO.   The resulting structural coordinates, reported in Table
\ref{tab:struct}, are in quite satisfactory agreement with experiment.
Our computed magnetic moments have magnitudes $3.4\,\mu_{\rm B}$
and rotate in the $b-c$ plane as one scans along $b$, and this
result hardly changes when SO is turned on.  Experimentally, the
magnetic moments on Mn$^{3+}$ sites are found to form an elliptical
spiral with amplitudes $m_b=3.9\,\mu_{\rm B}$ and $m_c=2.8\,\mu_{\rm B}$
\cite{Kenzelmann:2005}.  The origin of this ellipticity, which is
apparently not captured by our calculation, deserves further investigation.

\begin{table}[t]
\caption{\label{tab:struct} Experimental (Ref.~\onlinecite{alonso})
and theoretical (this work) structural parameters for orthorhombic
\tmo.}
\begin{ruledtabular}
\begin{tabular}{clccc}
 & & & Experiment & Theory\\
\hline
\multicolumn{2}{l}{Lattice vectors}
   & $a$ (\AA)                 & 5.293 & 5.195 \\
 & & $b$ (\AA)                 & 5.838 & 5.758 \\
 & & $c$ (\AA)                 & 7.403 & 7.308 \\
Tb & 4c($x$ $y$ 1/4) & $x$     & 0.983 & 0.979 \\
   &                 & $y$     & 0.082 & 0.084 \\
Mn & 4b(1/2 0 0)     & & & \\
O1 & 4c($x$ $y$ 1/4) & $x$     & 0.104 & 0.107 \\
   &                 & $y$     & 0.467 & 0.469 \\
O2 & 8d($x$ $y$ $z$) & $x$     & 0.704 & 0.699 \\
   &                 & $y$     & 0.326 & 0.320 \\
   &                 & $z$     & 0.051 & 0.052
\end{tabular}
\end{ruledtabular}
\end{table}

Next, we confirmed that the Berry-phase polarization vanishes in the
absence of SO.  When the SO is turned on,
one expects the broken inversion symmetry in the spin sector
to be communicated to the spatial (charge) degrees of freedom, and
indeed our calculations confirm this.  Focusing first on the purely
electronic mechanism, we froze the lattice vectors and ionic positions
while turning on the SOI and computed the polarization to be
$P=32\,\mu{\rm C}/{\rm m}^2$ along the $c$ direction\cite{uncert}
(and zero along directions $a$ and $b$).
This is much smaller than the experimental
value of $\sim600\,\mu{\rm C}/{\rm m}^2$ \cite{kimura}, suggesting that
the lattice mechanism must be dominant.  We thus turned our attention
next to the role of ionic displacements.

As a first step, we computed the Hellmann-Feynman forces appearing
on the ions as the SO is turned on (with the structural coordinates
still clamped).  Using the 20-atom cell of the $Pbnm$ space group to
label the forces, we found that:
(i) 64.8\% of the forces belong to the symmetry-preserving A$_1(\Gamma)$
irreducible representation (irrep);
(ii) 28.7\% of the forces belong to the B$_{\rm 3u}(\Gamma)$
irrep, i.e., zone-center infrared-active (IR-active) modes with
dynamical dipoles along $c$;
and (iii) 6.5\% correspond to forces at $k_y=\pm2\pi/3b$, i.e.,
those that would be responsible for lowering the translational
symmetry of the 20-atom cell.
The remaining contributions are found to vanish to
numerical precision.  Since the only forces
that generate an electric polarization in linear order are those
of type (ii), we henceforth focus our attention on the zone-center
IR-active B$_{\rm 3u}$ modes.

Eight Wyckoff coordinates contribute to the B$_{\rm 3u}$ irrep in the
20-atom {\it Pbnm} structure:
three associated with Mn atoms in Wyckoff position 4b,
one each for Tb and O1 atoms in Wyckoff position 4c,
and three for the O2 atoms in Wyckoff position 8d \cite{bcs:sam}.
To find the polarization induced by the SOI,
we need to deduce the displacements resulting from
these forces.  To this end we calculated the 8$\times$8
force-constant matrix $\Phi_{ij}=-\partial{f_i}/\partial{u_j}$
connecting the B$_{\rm 3u}$ forces and displacements at linear order using
finite-difference methods (i.e., we shift each Wyckoff coordinate
by $\sim$$10^{-3}$\,\AA\ and calculate the resulting forces).
For computational convenience, the determination of $\Phi_{ij}$ was
done in the absence of SO.  (We checked a few elements and found
hardly any difference in the presence of SO \cite{explan-so}.)
Furthermore, we identify the linear combination
$t_j$ of the eight coordinates corresponding to a uniform shift of the
crystal along the $c$ direction (acoustic mode) and make sure to
project this combination out of the computed forces and
displacements.  Thus, the displacements are predicted from
$u_i=-\sum^8_{j=1}\tilde\Phi^{-1}_{ij}f_j$,
where $\tilde\Phi^{-1}$ is a pseudo-inverse \cite{Xifan} having the
properties that
$\tilde\Phi^{-1} \Phi u=u$ for any $u$ orthogonal to $t$.

The computed forces and predicted atomic displacements in the
IR-active B$_{\rm 3u}$ sector for \tmo\ are presented in the first
three columns of Table \ref{tab:wyck_modes}.  We calculated the
B$_{\rm 3u}(\Gamma)$ forces again in the presence of the predicted
displacements and found that 99\% of these forces
were eliminated, thus justifying our approach
\cite{explan-fc}.  The distorted crystal structure is now
non-centrosymmetric (space group $Pna2_1$), and the
calculated Berry-phase polarization
(including both electronic and ionic contributions) is
$P=-467\,\mu{\rm C}/{\rm m}^2$.  This is more than an order of
magnitude larger than the value of 32\,$\mu$C/m$^2$ computed from
purely electronic mechanisms, and comparable in magnitude with the
experimental value of $\sim-$600$\,\mu$C/m$^2$ \cite{Yamasaki-both}.
Note that the sign of the ionic contribution is different from the 
purely electronic contribution, making the sign of the total polarization
consistent with experiment.
We thus arrive at the first major conclusion of our paper, namely,
that the lattice contributions dominate strongly
and have approximately the right magnitude
to explain the experimentally observed polarization.
Purely electronic mechanisms, e.g.\ as described by the
Katsura-Nagaosa-Balatsky (KNB) model\cite{Katsura:2005} or related
work \cite{Bruno:2005,jia:2006,jia:2007}, are not sufficient to
describe the ferroelectricity in \tmo.

\begin{table}[t]
\caption{\label{tab:wyck_modes} Forces, displacements, effective charges
and contributions to ionic polarization from IR-active B$_{\rm 3u}$ modes.}
\begin{ruledtabular}
\begin{tabular}{lrrrr}
~~Wyckoff        & $F$~~~ & $\Delta u$~~ & $Z^{*}$ & $\Delta{P}$~~~ \\
coordinate      & (meV/\AA) & (m\AA) & ($e$) & ($\mu$C/m$^2$) \\
\hline
 1 (Tb 4c, $z$) &  0.434 & $-$0.171 &  7.468 &  $-$94 \\
 2 (O1 4c, $z$) &  2.264 & $-$0.162 & $-$6.815 &   81 \\
 3 (Mn 4b, $x$) & $-$7.039 & $-$0.315 &  0.567 &  $-$13 \\
 4 (Mn 4b, $z$) & $-$8.933 & $-$0.453 &  7.460 & $-$248 \\
 5 (Mn 4b, $y$) & $-$2.940 & $-$0.012 &  0.553 &   $-$0.5 \\
 6 (O2 8d, $x$) &  5.058 &  0.540 &  0.077 &    3 \\
 7 (O2 8d, $y$) &  3.573 &  0.934 &  0.227 &   16 \\
 8 (O2 8d, $z$) &  4.409 &  0.556 & $-$5.738 & $-$234
\end{tabular}
\end{ruledtabular}
\end{table}

We focus henceforth on the lattice-mediated contribution to the
polarization.
To understand which modes contribute most strongly and
to test whether simple models (based, e.g., on DM interactions)
can predict the forces and displacements, we performed
a further analysis as follows.  For each
B$_{\rm 3u}$ Wyckoff coordinate, we computed the corresponding effective
charge $Z^*$ (derivative of polarization with respect to the Wyckoff
coordinate) by augmenting the same finite-difference calculations
used earlier with Berry-phase polarization calculations.  We then
multiplied effective charges and displacements to get the contribution
$\Delta P$ made by each Wyckoff coordinate.
The results are given in the last two columns of Table \ref{tab:wyck_modes}.
The total ionic polarization (sum of the last column) is
$-$489\,$\mu$C/m$^2$, in good agreement with the change of polarization
in going from the centrosymmetric to the distorted structure as
reported earlier ($-$467$-$32=$-$499\,$\mu$C/m$^2$).
Not surprisingly, the $Z^*$ values are much larger for the four Wyckoff
coordinates involving displacements along $z$, and the corresponding
$\Delta P$ values give by far the largest contribution.

In a simple DM model \cite{Sergienko:2006a}, one considers interactions
involving nearest-neighbor triplets of ions along Mn--O--Mn bonds.
Such a picture leads to the expectation that a force
$\gamma\,\hat{{\bf e}}_{nn'}\times {\bf S}_n\times {\bf S}_{n'}$ 
should appear on the central
O ion (where $\gamma$ is a coupling, ${\bf S}_n$ and ${\bf S}_{n'}$ are
neighboring Mn spins, and
$\hat{{\bf e}}_{nn'}$ is the unit vector connecting them),
with an equal and opposite force shared between the Mn ions.
The cycloidal spin structure of \tmo\ is such that O1 ions should be
unaffected, and O2 ions should all feel the same force along
$c$, with Mn ions feeling an opposite force.  Thus, within
such a model one would expect only the 4th and 8th rows of
Table \ref{tab:wyck_modes} to give significant contributions.
Inspecting the Table, we see that
this is rather far from being the case; one does find substantial
oppositely-directed forces and displacements for the (Mn 4b, $z$)
and (O2 8d, $z$) coordinates, but the forces and displacements
are comparable for some other coordinates.  The
contributions to $\Delta P$ are more strongly
dominated by these two coordinates, but only because
the $Z^*$ values are so much larger for displacements
along $\hat{z}$.
Interestingly, the Tb displacements are responsible for $\sim$20\%
of the lattice polarization, despite being uninvolved
in the usual picture of Mn--O--Mn couplings.
Thus we conclude that the simple DM model, and
indeed any model that considers only nearest-neighbor Mn--Mn spin
interactions, is unable to provide a detailed description
of the ferroelectricity in \tmo.

\begin{table}[t]
\caption{\label{tab:eig_modes} Forces, displacements, effective charges
and contributions to ionic polarization from modes corresponding
to eigenvectors of the force-constant matrix.}
\begin{ruledtabular}
\begin{tabular}{rrrrr}
Eigenvalue   & $F$~~~    & $\Delta{u}$~~ & $Z^{*}$ & $\Delta{P}$~~~ \\
(eV/\AA$^2$) & (meV/\AA) & (m\AA)        & (e)       & ($\mu$C/m$^2$) \\
\hline
    34.86 &  6.349 &  0.182 & $-$8.111 & $-$108 \\
    31.37 & $-$8.259 & $-$0.263 &  4.565 &  $-$88 \\
    13.32 &  2.783 &  0.209 &  8.132 &  125 \\
    11.87 &  7.614 &  0.641 & $-$2.127 & $-$100 \\
    10.35 &  1.684 &  0.163 & $-$3.406 &  $-$41 \\
     6.99 & $-$2.784 & $-$0.398 &  4.316 & $-$126 \\
     3.85 &  4.032 &  1.047 & $-$1.991 & $-$153
\end{tabular}
\end{ruledtabular}
\end{table}

As an alternative way of analyzing the SO-induced B$_{\rm 3u}$
forces and displacements, we transform into the basis of
eigenvectors of the 8$\times$8 force-constant matrix $\Phi$.  The
results are presented in Table \ref{tab:eig_modes}.
(Components on the translational zero mode, not shown in
the Table, were numerically negligible.)
We find that the forces are rather widely distributed among these
``modes.''  Not surprisingly, the displacements are largest for
the softest mode, but other modes also show substantial displacements.
Moreover, the mode dynamical charges are much larger for some of
the harder modes, with the result that the contributions to
the polarization are of comparable magnitude for almost all of the
B$_{\rm 3u}$ modes.  Thus, it is clear that the physics of the
electric polarization in TbMnO$_3$ is not dominated by a single
softest mode.

To investigate the role of spin-orbit effects located on the various atomic
sites, we performed additional calculations of the
forces (and, via the force-constant matrix, the resulting
lattice polarizations).  We first turned
the SOI off on all sites other than Tb, then Mn, then
O, finding contributions to the polarization of
$-11\,\mu$C/m$^2$, $-447\,\mu$C/m$^2$ and $-8\,\mu$C/m$^2$,
respectively.  (The sum is in fairly good agreement with the total
ionic polarization of $-$489\,$\mu$C/m$^2$; this and other tests
confirm that the results are roughly linear in SO coupling.)
This confirms that the SOI on the Mn sites is responsible for almost
all of the lattice-mediated contribution to the polarization, even
though some of these forces appear on other sites,
even including Tb (see Table \ref{tab:wyck_modes}).

Up to now our analysis has been done using the 60-atom
supercell, whose cycloidal wavevector ($k_{\rm s}$=1/3)
approximates the experimental one ($k_{\rm s}$=$0.28$).
To investigate how the results depend
on $k_s$, we have also carried calculations of the SO-induced
forces for a 40-atom supercell ($k_{\rm s}$=$1/2$).  In order to
isolate the effects of the spin-wave period, we calculated the
forces at the same strucural coordinates as for the 60-atom
structure, and again projected into the B$_{\rm 3u}(\Gamma)$ irrep.
We find that the pattern of B$_{\rm 3u}$ forces matches almost
exactly that of the 60-atom cell (with the angle between the two
eight-dimensional vectors being $\sim$3$^\circ$), although the
magnitude is larger for the 40-atom cell.  From the simple DM model
one expects
${\bf P} \propto {\bf e}_{n,n'}\times({\bf S}_n\times{\bf S}_{n'})$
\cite{Kimura:2007} where the spins are neighbors on the zig-zag chains
propagating along $b$ (see Fig.~\ref{fig:rums}).  The angle between
the spins is 90$^\circ$ and 60$^\circ$ for the 40-atom and 60-atom
supercells respectively, giving an expected ratio
of $2:\sqrt{3}$=1.15.  In the long-wavelength
limit one expects ${\bf P} \propto |k_{\rm s}|$, for a ratio of 1.5.
Strangely, our computed ratio for the 40- and 60-atom supercells is
about 1.48,
closer to the continuum expression in spite of the rather short
wavelength of the spiral.  This may be
another indication that further-neighbor Mn--Mn
interactions, or other effects not captured by a
nearest-neighbor DM model, play an important role.
From the forces we predict a lattice
polarization of $-754\,\mu$C/m$^2$ for $k_{\rm s}$=1/2;
extrapolating through our $k_{\rm s}$=1/3 result, we estimate a
lattice polarization of $-410\,\mu$C/m$^2$ for $k_{\rm s}$=0.28.

In summary, we have used first-principles methods to compute the
electronic and lattice contributions to the spin-orbit induced
electric polarization in the cycloidal-spin compound \tmo, and
find the lattice contribution to be strongly dominant.
We provide a detailed analysis of the lattice contributions coming from
individual sites and individual modes and roughly characterize
the dependence of the lattice polarization on spin-wave period.
Our result for the polarization agrees in sign and compares fairly
well in magnitude with the experimentally measured value of 
$\sim-$600$\,\mu$C/m$^2$ \cite{Yamasaki-both}.

The authors wish to thank Morrel H.\ Cohen and Craig J.\ Fennie
for fruitful discussions.
This work was supported by NSF Grant No. DMR-0549198.

{\it Note:} While finalizing this manuscript, we learned of
closely related work by H.J.\ Xiang {\it et al.} \cite{Xiang:2008}

\bibliography{tmo}

\begin{thebibliography}{41}
\bibitem[{unc()}]{uncert}
\bibinfo{note}{The uncertainty in the calculated values of polarization is
  estimated to be of the order of 5-10$\,\mu{\rm C}/{\rm m}^2$.}
\bibitem[{exp({\natexlab{a}})}]{explan-so}
\bibinfo{note}{In any case, a dependence of $\Phi_{ij}$ on SOI would only give
  a dependence of polarization at quadratic or higher order. We are primarily
  concerned with linear effects.}
\bibitem[{exp({\natexlab{b}})}]{explan-fc}
\bibinfo{note}{One could alternatively just relax the atomic coordinates in the
  presence of SOI. However, we prefer the present approach because it is
  numerically more stable, it facilitates analysis of distortions and
  polarizations in terms of phonon modes, and it allows for efficient testing
  of various effects (e.g., turning off the SOI on certain sites) without the
  need to perform additional ionic relaxations.}
\end{thebibliography}

\end{document}